\documentclass[12pt]{article}
\usepackage{graphicx}
\begin{document}

\pagestyle{empty}

\noindent
{\bf ASAS-SN Observations of the Pulsation of some R Coronae Borealis (RCB) Stars}

\bigskip

\noindent
{\bf John R. Percy\\Department of Astronomy and Astrophysics, and\\Dunlap Institute of Astronomy and Astrophysics\\University of Toronto\\Toronto ON\\Canada M5S 3H4\\john.percy@utoronto.ca}

\medskip

\bigskip

{\bf Abstract}  Photometry from the All-Sky Automated Survey for Supernovae
(ASAS-SN), along with the AAVSO VSTAR time-series analysis package has been
used to study the pulsational properties of 31 R Coronae Borealis (RCB)
stars.  Periods have been derived for many of the stars, but the variability
tends to be small and complex, occasionally multiperiodic, with noticeable variation in amplitude.
As with other RCB stars, the periods are a few weeks.

\medskip


\medskip

\noindent
ADS keywords = stars; stars: chemically peculiar; techniques: photometric; methods: statistical; stars: variable; stars: oscillations

\medskip

\noindent
{\bf 1. Introduction}

\smallskip

R Coronae Borealis (RCB) stars are rare carbon-rich, hydrogen-poor, highly-evolved yellow supergiants
which undergo fadings of up to 10 magnitudes, then slowly return to normal
(maximum) brightness; see Clayton (2012) for an excellent review.  
Most or all RCB stars also undergo small-amplitude pulsations with
periods of a few weeks; see Rao and Lambert (2015), hereafter RL, for a discussion and list and some references.  

Although it was once thought that the fadings of RCB stars
were random, it is now known that, in at least some of them, the fadings
are locked to the pulsation period i.e. the onsets of the fadings occur at about
the same phase of the pulsation cycle (Pugach 1977, Lawson {\it et al.} 1992, Crause {\it et al.} 2007).  This suggests a causal connection: e.g. the pulsation ejects a cloud of gas and
dust; when this cools, the carbon condenses into soot; if the cloud lies
between the observer and the star, the star appears to fade; it slowly
reappears as the cloud disperses.  It is also possible that temperature and
density fluctuations in the stellar atmosphere, during the pulsations, lead
to dust condensation (e.g. Woitke {\it et al.} 1996).  Either case
implies that the mass ejection is not
radially symmetric; a cloud is ejected, not a shell.

Percy and Dembski (2018) have recently published an analysis of the pulsation
of the bright RCB star RY Sgr during a long interval at maximum light.  They
found significant variations in amplitude, and ``wandering" of the period.  They
pointed out that, for the analysis of other RCB stars, long intervals of
precise photometry at maximum were necessary.  ASAS-SN data might well be suitable. 
The present paper presents an
analysis of the pulsation of 31 RCB stars, using data from that survey.
 
\medskip

\noindent
{\bf 2. Data and Analysis}

\smallskip

The All-Sky Automated Survey for Supernovae (ASAS-SN) uses a network
of up to 24 telescopes around the world to survey the entire visible sky
every night down to about 18th magnitude, and has been doing so for over
2000 days (Shappee {\it et al.} (2014), 
Jayasinghe {\it et al.} (2018)).   The latter paper presents a uniform
classification of 412,000 known variable stars -- a remarkable contribution.
 
The data can be accessed in one of two ways.  (1) ASAS-SN has
identified over 50,000 variable stars, classified these, determined
periods and amplitudes for those that are periodic, and made this information
publicly available on-line, along with the photometry: asas-sn.osu.edu/variables.  (2) For
stars not listed in the ASAS-SN variable star catalogue, ASAS-SN images
can be queried and measured, at any given position on the sky (a time-consuming process!): asas-sn.osu.edu.  In either
case, the photometric measurements can then be downloaded for analysis.  ASAS-SN also
provides the mean V magnitude, magnitudes in several other filters, the (J-K) color, the
parallax and distance.  For some complex stars such as pulsating red giants, the ASAS-SN 
classifications and periods can be problematic.  We shall address that 
problem in a subsequent paper.
For a few stars, we also analyzed visual and/or PEP observations from the
AAVSO International Database (AID: Kafka 2019).

The data were analysed with the AAVSO VSTAR time-series analysis package
(Benn 2013), to display the light curve, to carry out Fourier analysis and,
for a few stars, to investigate possible amplitude and/or period variation
using wavelet analysis.  This package is publicly-available on the AAVSO website.

\medskip

\noindent
{\bf 3 Results}

\smallskip

We analyzed 31 of the RCB stars in Table 1 of Clayton (2012).  During
the time interval covered by the ASAS-SN data, many of the other stars
were undergoing fadings, and showing complex non-pulsational variability for that reason.
Others were too faint for the ASAS-SN measurements to be useful; many
measurements were upper limits.  For
others, the scatter was large, and no periodic signals stood out above
the noise.  Our sample is therefore a restricted and slightly-biased one.  The 31 
stars in Table 1 yielded some results, and notes on these are given
below. The Table includes the star name, the maximum V magnitude (from Clayton (2012)), the period PRL in days given by RL, and the period in days and amplitude obtained in the present study.
 Of the 31 stars, 13 were
listed by RL.  In the last column, v 
indicates that the amplitude was significantly
variable, and a colon (:) indicates uncertainty.  Light curves of two representative stars are
shown in Figures 1 and 2.

\medskip

\begin{figure}
\begin{center}
\includegraphics[height=9cm]{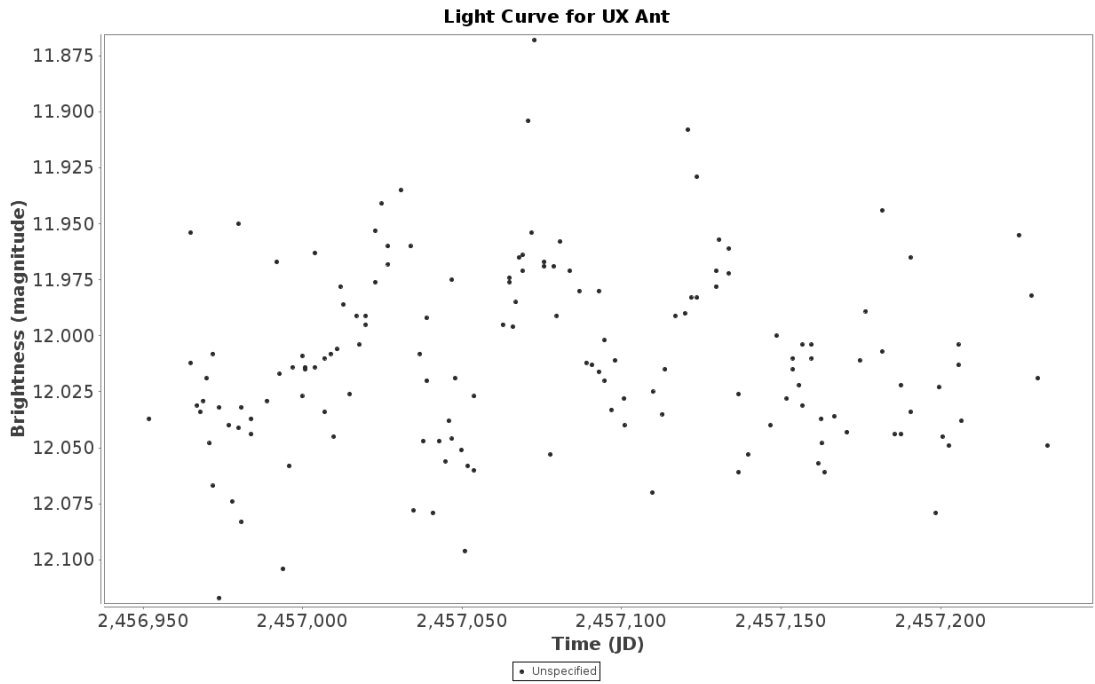}
\caption{The ASAS-SN V light curve of UX Ant, showing cycles of increase and
decrease on a time scale of about 50 days, with an average (but variable)
peak-to-peak range of 0.1 mag.} 
\end{center}
\end{figure}

\begin{figure}
\begin{center}
\includegraphics[height=9cm]{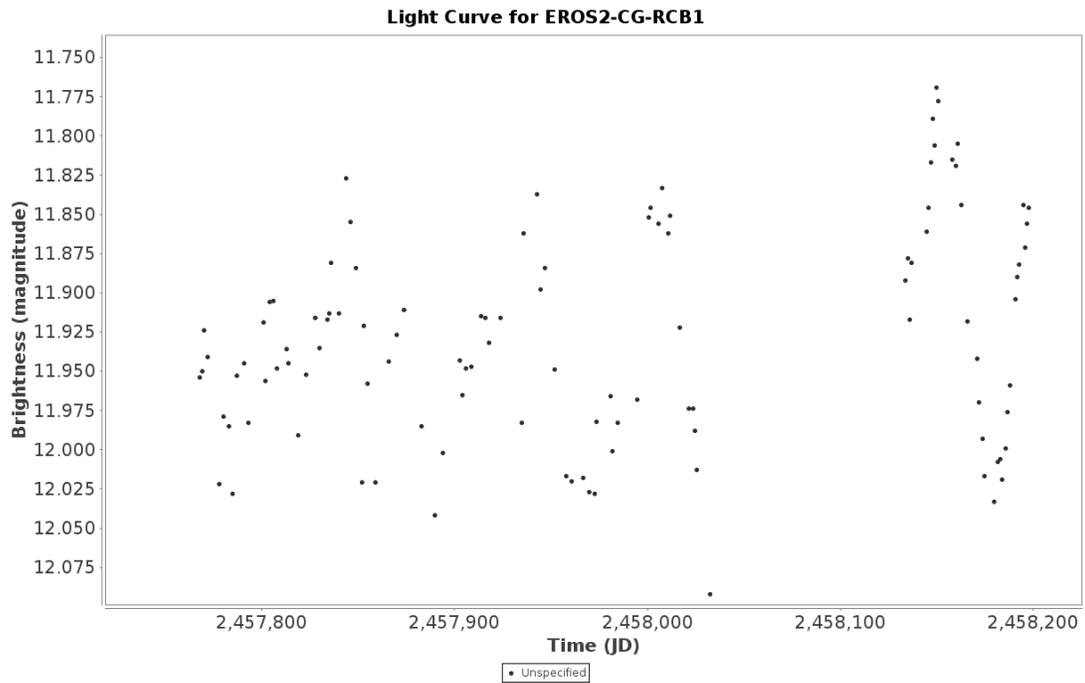}
\caption{The ASAS-SN V light curve of EROS2-CG-RCB1 (V409 Nor), showing cycles
of increase and decrease on a time scale of about 50 days.}
\end{center}
\end{figure}

\begin{figure}
\begin{center}
\includegraphics[height=9cm]{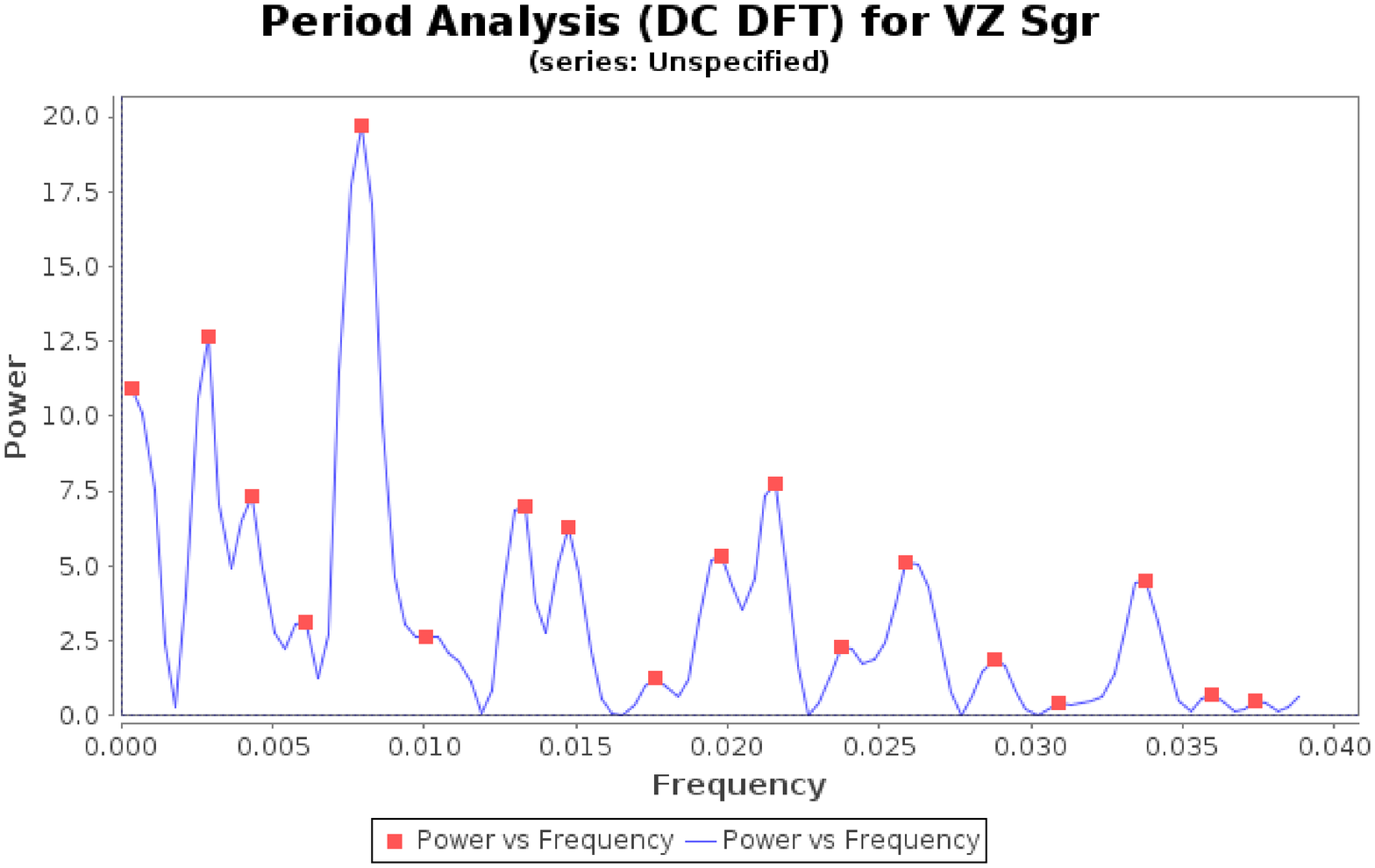}
\caption{The Fourier spectrum of the ASAS-SN V light curve of VZ Sgr, showing
a conspicuous peak at a frequency corresponding to a period of 126 days.
This period is longer than the pulsation periods of most RCB stars, but not
uniquely so (Table 1).}
\end{center}
\end{figure}

\noindent
{\bf 3.1 Results on Individual Stars}

\smallskip

{\it XX Cam:}  There is a conspicuous peak at a period of 29.5 days,
and an
amplitude of 0.04 mag, but the amplitude is highly variable.  In the light curve,
the cycle lengths are about 30 days.  AAVSO V data, however, show strongest
peaks at 19 to 24 days, but with small amplitude.  RL give a period of 36 days.

{\it UX Ant:}  There are conspicuous peaks at periods of 57.9 days and 48.8
days; these are aliases.  The former period has a slightly larger amplitude.
Figure 1 shows the light curve.  RL give a period of 50 days.

{\it UW Cen:}  The highest peak is at 43.5 days, but it is not conspicuous.
On the other hand, RL give a period of 42.8 days, which is very close.  The
light curves are scattered.

{\it Y Mus:}  There are comparable peaks at 34.7 and 38.41 days; these are aliases.  The light curves are scattered.  RL give a period of 35.0 days.

{\it DY Cen:}  The highest peak is at a period of 29.87 days, with an 
amplitude of 0.02 mag.  The highest peak in the AAVSO visual data is at a
period of 29.49 days.

{\it V854 Cen:}  For some reason, there is scatter of over a magnitude in
the ASAS-SN data.  RL give a period of 43.25 days.

{\it S Aps:}  The highest peak is at a period of 123 days, with an amplitude
of 0.055 mag., but there may be smaller, shorter cycles also.  The 123-day
period is consistent with the period of 120 days given by RL.

{\it ASAS-RCB-1 (V409 Nor):}  The highest peak is at a period of 49.9 days, with an
amplitude of 0.047 mag, but there are other peaks which are almost as high.
Cycle lengths are 50-55 days.  Figure 2 shows the light curve.


{\it RT Nor:}  The highest peaks in the
spectrum of the ASAS-SN data are at 43.44 and 58.8 days.  The cycle lengths
are about 56 days.  The highest peaks
in the spectrum of the AAVSO visual data is at 55.49 days with an amplitude
of 0.036.  At this point then, we cannot choose between 43.44 and 58.8 days.
RL give a period of 43 days.

{\it ASAS-RCB-3:}  The highest peak in the spectrum of the ASAS-SN data is
at 47.28 days, with an amplitude of 0.049, though this peak does not
stand out strongly.

{\it V517 Oph:}  The highest peak in the spectrum of the ASAS-SN data is at
27.5 $\pm$ 1 days.  This star was emerging from a fading.

{\it ASAS-RCB-10:}  The highest peak in the spectrum of the ASAS-SN data is
at 58.63 days, with an amplitude of only 0.021.

{\it V2552 Oph:}  The highest peak in the spectrum of the ASAS-SN data is at
48.71 days, with an amplitude of 0.026, but this peak does not stand out
strongly.

{\it EROS2-CG-RCB5:}  The highest peak in the spectrum of the ASAS-SN data is
at 29.56 days, with an amplitude of 0.040.


{\it EROS2-CG-RCB-1:}  The pulsation amplitude, if any, is less than 0.05 mag,
but there appear to be cycles of both 35-50 days and greater than 100 days in the light curve.

{\it EROS2-CG-RCB-2:}  Very faint.  The highest peak in the spectrum is at
44.44 days, with an amplitude of 0.11 mag, and the light curve shows cycles
of this order.

{\it V739 Sgr:}  During an approximately one-year interval when it was at
maximum, the highest peak in the spectrum was at 45.8 days, with an amplitude
of 0.07 mag. 

{\it EROS2-CG-RCB-14:}  The highest peak in the spectrum is at 27.45 days, and
there is some evidence in the light curve for cycles on this time scale, but
the amplitude is less than 0.05 mag in peak-to-peak range.

{\it V3795 Sgr:}  The highest peak in the spectrum is at about 35$\pm$3 days,
and the light curve shows cycles of this order.

{\it VZ Sgr:}  The highest peak is at 126.4 days, with an amplitude of 0.07 mag (Figure 3),
and there are cycles in the light curve of this time scale.
The ASAS-SN website gives a period of 129 days, and the AAVSO
visual data are consistent with our value.  RL give a period of 40 days.

{\it IRAS 18135-2419:}  The highest peak in the spectrum is at 132.08 days,
with an amplitude of 0.07 mag; the second-highest peak is half of that.  The cycles in the light curve are consistent with the
longer period. 


{\it GU Sgr:}  The light curve is dense and complex.  The highest peak in
the spectrum is at 35.00 days, and the cycles in the light curve are entirely consistent
with this period, though wavelet analysis shows that the amplitude is variable by more than a factor of three, from 0.04 to 0.14 mag.  The highest
peak in the spectrum of the AAVSO visual data is at 57.54 days, though the
second-highest is at 36.64 days.

{\it V391 Sct:}  Cycle lengths in the light curve are about 50 days.

{\it NSV 11154:}  This star is undergoing a complex mixture of variations,
but intervals of the ASAS-SN data, and AAVSO V data give periods in the
range 45 to 48.5 days, with mean amplitudes of 0.1 mag.

{\it MV Sgr:}  There is a fading during the interval of observation but, in
the short interval JD less than 2457300 when the star is at maximum, there
is a peak at 30.06 days with an amplitude of 0.02 mag.

{\it FH Sct:}  The highest peaks in the spectrum are at 43$\pm$5 days,
with an amplitude of about 0.08 mag.  The AAVSO V data shows cycle lengths
of about 50 days.  RL give a period of 41 days.

{\it ASAS-RCB-8:}  There are several peaks in the range of 50 to 65 days, with
amplitudes of 0.05 mag, though the variation is quite irregular.

{\it SV Sge:}  The strongest peak in the spectrum is at 60.94 days, with an
amplitude of 0.04 mag.

{\it RY Sgr:}  The highest peak in the spectrum is at 38 $\pm$ 1 days,
with an amplitude of 0.17 mag, but the period and amplitude are both variable.
The pulsation of this star has been well-studied; see Percy and Dembski (2018)
for references.  RL give a period of 37.79 days.

{\it U Aqr:}  There are peaks in the spectrum at 34, 37 and 78 days, all with
amplitudes of about 0.1 mag, and there are cycles in the light curve with
both these time scales, and with quite large amplitudes. 
RL give a period of 81.3 days, consistent with the
present results.  This object has, in the past, been suggested as a possible
Thorne-Zytkow object (Vanture {\it et al.} 1999).

{\it UV Cas:}  The highest peak in the spectrum is at 65.04 days, with an
amplitude of only 0.037 mag, but there are other peaks which are almost
as high.  RL give a period of 40.16 days, and AAVSO visual and V data also give a period
of 41 to 42 days, but with a small amplitude.

\medskip

\noindent
{\bf 4. Discussion}

\smallskip

The results of this study are limited by the fact that the pulsation
is complex (possibly multiperiodic in some stars), its amplitude is small and variable,
many of the stars are faint (and the errors of the photometry are
significant), and the dataset is only 2000 days long, and affected by
seasonal gaps.

For three stars -- XX Cam, DY Cen, and EROS2-CG-RCB5 -- we obtain a period
close to one synodic month.  It is not clear why there should be a spurious 
period at one synodic month.  Furthermore, 29.5 days is a quite reasonable
pulsation period for an RCB star.

The ASAS-SN photometry database is obviously an excellent resource, with many
applications.  Many of the new variables in the database have not been fully
analyzed, especially if the variability is complex and/or apparently not
periodic.  For some purposes, the ASAS-SN V data can be combined with other
data, such as that in the AAVSO International Database, to get a more complete
understanding of the variability.  For purposes of, for example, long-term
period analysis of individual pulsating or eclipsing variables, the ASAS-SN photometry
can add a few more points to the (O-C) diagram.

The ASAS-SN database has obvious educational applications.  The analysis of
variable star data enables students to develop and integrate their skills in
math, science, and computing, motivated by doing real science with real
data (e.g. Percy 2018).  There are more than enough ``unsolved" or under-analyzed variables
in the ASAS-SN database to keep students busy.  The AAVSO VSTAR package is
well-suited for analysis by students, and others.  I anticipate a series of
future papers, showing some examples from my own students' work.  Likewise,
the ASAS-SN database, along with VSTAR, also provides opportunities for
amateur (and professional) astronomers who are interested in variable star analysis as a
cloudy-night activity.

\medskip

\noindent
{\bf 5. Conclusions}

\smallskip

Almost all of the RCB stars that were analyzed showed some evidence for
small-amplitude pulsational variability on time scales of weeks.  Most
of the periods were between 20 and 60 days, but a handful of stars showed
credible periods of up to 130 days.  The variability
was most often complex, with variable amplitude and, in a few cases,
evidence for multiperiodicity.  For other stars in Clayton (2012), not listed in Table 1,
the data were too sparse and/or noisy to provide information on pulsational
variability.

\medskip

\noindent
{\bf Acknowledgements}

\smallskip

The author thanks the ASAS-SN team for making the photometry publicly
available, in user-friendly form, and the developers of VSTAR for creating
the package and making it easily available and user-friendly.
This project made
use of the SIMBAD database, maintained in Strasbourg, France.
The Dunlap Institute is funded through an endowment established by the David Dunlap family and the University of Toronto.

\bigskip

\noindent
{\bf References}

\smallskip

\noindent
Benn, D. 2013, VSTAR data analysis software {http://www.aavso.org/vstar-overview).

\noindent
Clayton, G.C. 2012, {\it J. Amer. Assoc. Var. Star Obs.}, {\bf 40}, 539.

\noindent
Crause, L.A., Lawson, W.A., and Henden, A.A. 2007, {\it Mon. Not. Roy. Astron.
Soc.}, {\bf 375}, 301.

\noindent
Jayasinghe, T. {\it et al.} 2018, {\it arXiv:} 1809.07329.

\noindent
Kafka, S. 2019, variable star observations from the AAVSO International
Database 

(https://www.aavso.org/aavso-international-database)

\noindent
Lawson, W.A., Cottrell, P.L., Gilmore, A.C., and Kilmartin, P.M. 1992,
{\it Mon. Not. Roy. Astron. Soc.}, {\bf 256}, 339.

\noindent
Percy, J.R. 2018, in {\it Robotic Telescopes, Student Research and Education},
Ed. M. Fitzgerald {\it et al}., Vol. 1, No. 1, 95.

\noindent
Percy, J.R. and Dembski, K.H. 2018, {\it J. Amer. Assoc. Var. Star. Obs.}, {\bf 46}, 127.

\noindent
Pugach, A.F. 1977, {\it Inf. Bull. Var. Stars}, No. 1277, 1.

\noindent
Rao, N.K., and Lambert, D.L. 2015, {\it Mon. Not. Roy. Astron. Soc.}, {\bf 447}, 3664.

\noindent
Shappee, B.J. {\it et al}. 2014, {\it Astrophys. J.}, {\bf 788}, 48S.

\noindent
Vanture, A.D., Zucker, D., and Wallerstein, G. 1999, {\it Astrophys. J.}, {\bf 514}, 932.

\noindent
Woitke, P., Goeres, A., and Sedlmayr, E. 1996, {\it Astron. Astrophys.},
{\bf 313}, 217.

\medskip

\begin{table}
\begin{center}
\caption{Pulsation Properties of RCB Stars from ASAS-SN V Photometry}
\begin{tabular}{rrrrl}
\hline
Star & Vmax & PRL (d) & P (d) & V Amp \\
\hline
XX Cam & 8.7 & 36 & 29.5 & 0.04v \\
UX Ant & 12.2 & 50 & 49 or 58 & 0.03v \\
UW Cen & 9.6 & 42.79 & 43.5 & 0.06 \\
Y Mus & 10.5 & 35.0 & 35 or 38 & 0.04 \\
DY Cen & 12.0 & -- & 29.5-29.9 & 0.04 \\
V854 Cen & 7.0 & 43.25 & -- &  0.1: \\
S Aps & 9.6 & 120 & 123 & 0.06 \\
ASAS-RCB-1 & 11.9 & -- & 49.9 & 0.05v \\
RT Nor & 11.3 & 43 & 43 or 59 & 0.07v \\
ASAS-RCB-3 & 11.8 & -- & 47.28 & 0.05 \\
V517 Oph & 12.6 & -- & 27.5 & 0.05: \\
ASAS-RCB-10 & 11.5 & -- & 58.63 & 0.02: \\
V2552 Oph & 10.8 & -- & 48.71: & 0.03 \\
EROS2-CG-RCB-5 & 13.5 & -- & 29.56 & 0.04 \\
EROS2-CG-RCB-1 & 14.6 & -- & 35-50 & $\le$0.05 \\
EROS2-CG-RCB-2 & 14.5 & -- & 44.44 & 0.1: \\
V739 Sgr & 14.0 & -- & 45.8 & 0.06 \\
EROS2-CG-RCB-14 & 12.5 & -- & 27.45 & 0.03: \\
V3795 Sgr & 11.5 & -- & 35$\pm$3 & 0.05: \\
VZ Sgr & 11.8 & 40 & 126-9 & 0.07 \\
IRAS 18135-2419 & 12.8 & -- & 132.08 & 0.07 \\
GU Sgr & 11.3 & -- & 35.00 & 0.1v \\
V391 Sct & 13.3 & -- & 50 & 0.04: \\
NSV 11154 & 12.0 & -- & 45-9 & 0.1v \\
MV Sgr & 12.0 & -- & 30.06 & 0.02 \\
FH Sct & 13.4 & 41 & 43$\pm$5 & 0.08 \\
ASAS-RCB-8 & 10.9 & -- & 50-65 & 0.05v \\
SV Sge & 11.5 & -- & 60.94 & 0.05v \\
RY Sgr & 6.5 & 37.79 & 38$\pm$1 & 0.17v \\
U Aqr & 10.5 & 81.3 & 34.37 or 78 & 0.10 \\
UV Cas & 11.8 & 40.16 & 65.04 or 41-42 & 0.04 \\
\end{tabular}
\end{center}
\end{table}

\end{document}